 \documentclass[final,5p,times,twocolumn]{elsarticle}

\usepackage{graphicx}
\usepackage{amssymb}
\usepackage{epsfig} 
\usepackage{epstopdf}

\newcommand{\q}{$q_{\rm n}\,$}

\journal{Nuclear Instruments and Methods A}

\begin{document}

\begin{frontmatter}



\title{An optical device for ultra-cold neutrons -\\ 
Investigation of systematic effects and applications}


\author[1]		{C. Plonka-Spehr}
\ead{plonka@uni-mainz.de}
\author[1]		{A. Kraft}
\author[2,3]	{P. Iaydjiev}
\author[4]		{J. Klepp}
\author[2]		{V. Nesvizhevsky}
\author[2]		{P. Geltenbort}
\author[1]		{Th. Lauer}


\address[1]{Institute for Nuclear Chemistry, University of Mainz, Germany}
\address[2]{Institut Laue-Langevin, Grenoble, France}
\address[3]{Institute for Nuclear Research and Nuclear Energy, Sofia, Bulgary}
\address[4]{Faculty of Physics, University of Vienna, Austria}

\begin{abstract}
We developed an optical device for ultra-cold neutrons and investigated the influence of a tilt
of its guiding components. A measurement of the time-of-flight of the neutrons through the device by means of a dedicated chopper system was performed
and a light-optical method for the alignment of the guiding components is demonstrated.
A comparative analysis of former experiments with our results shows the potential of such a device
to test the electrical neutrality of the free neutron on 
the $10^{-22}q_{\rm e}$ level and to investigate the interaction of neutrons with gravity.
\end{abstract}


\end{frontmatter}


\section{Introduction}
\label{sec:Introduction}
Electric charge quantization (ECQ) and atom neutrality are 
well-established experimental observations. Their understanding is a long standing
question of basic interest. 
ECQ implies that the charges of all known
particles can be derived from integer multiples of one fundamental charge namely that of the electron.
Early attempts to explain ECQ in the framework of quantum electrodynamics, like Dirac's  assumption of magnetic monopoles \cite{Dirac1931}, still lack in experimental evidence. 

In today's framework of the Standard Model (SM), the quantization of charge is associated with 
the classical structure of the theory (i.e. gauge invariance) and the cancellation of quantum anomalies. 
As stated in \cite{foot-1993-19,PhysRevD.49.3617,PhysRevD.51.2411}, 
ECQ might not be a natural consequence of this model but can be effectuated with additional constraints
to the SM Lagrangian by adding right-handed neutrino masses with Dirac and Majorana terms.
The theoretical situation is ambiguous however, as the presence of these terms could also shift the electric charges 
of the known particles by an amount proportional to their baryon-lepton $(B-L)$ number \cite{arvanitaki:120407}. 
As a consequence, so called ÔneutralÕ particles like the neutron could carry a small 'rest charge'. 

In superstring models or grand unified groups, ECQ is a natural result of the 
theory \cite{Superstring1986,GUT1985}. But also conventional 
extensions to the SM could lead to ECQ with the appearance of new fermions  \cite{PhysRevD.49.3617}. 
On the contrary, other hypotheses concern a new class of SU(2) $\times$
U(1) models, where dequantization of the electric charge can occur by a nonzero photon mass 
\cite{Ignatiev1996} or by mini-charged Higgs bosons which couple to fermions \cite{PhysRevD.44.3706}.  
In \cite{PhysRevD.48.4481} it is suggested, that the charges of quarks and leptons, 
including neutrinos, and also of the proton and the neutron could have changed with the evolution of the Universe 
in time to a resulting present finite rest charge of the neutron or the neutrino.
Assumptions beyond the conventional framework suggest topological shifts of the SM electric  charges, where
ordinary particles carry quantum hairs under massive higher spin fields (see \cite{arvanitaki:120407} and references therein)
resulting in charged atoms and neutrons.

We recall from \cite{Feinberg1959} that a non-zero neutron charge would not be inconsistent with the 
conservation of the electric charge, the baryon and the lepton number. It would rather imply that conservation of 
the baryon number and conservation of the electric charge are not independent fundamental symmetries. 
Assuming charge conservation, a non zero value of  $q_{\rm n}$ would eliminate the possibility of neutron-antineutron oscillations \cite{Glashow1979}, 
which is a candidate for a violation of the baryon number by $\Delta B=2$. 

Today's accuracy on atom neutrality has reached a sensitivity of $10^{-21}\,q_{\rm e}$ \cite{Metrologia2004}
and hence is comparable to direct limits on the neutron charge $q_{\rm n}$ \cite{PhysRevD.37.3107}.
However, in \cite{PhysRev.153.1415} it is pointed out, that the charges of bound and free neutrons might not be identical. This would constrict results on the upper limit of the neutron charge obtained from indirect 
measurements on unionized atoms and molecules. 

\section{Direct measurements of $q_{\rm n}$}
Experiments searching for an electric charge of the free neutron were carried out shortly after neutron beams
were available from reactor sources.  
Shapiro and Estulin tried to deflect neutrons in a strong electric field \cite{Shapiro1956} and reached
a sensitivity of  $\delta q_{\rm n}<10^{-12}\,q_{\rm e}$. 
A significant improvement of this approach was achieved by means of neutron optical elements with crystals or gratings in 
order to detect shifts of the beam profile or its angular deviation. 
Shull et al.\,\cite{PhysRev.153.1415} obtained
a value for the ratio of the neutron over the proton charge, using a double crystal spectrometer, as
$(-1.9\pm 3.7)\times10^{-18}$. 

\section{Present experimental situation}
The most precise direct limit on the free neutron charge was achieved 20 years ago
with a setup described in \cite{PhysRevD.37.3107} using a beam of cold neutrons
at the Institut Laue-Langevin (ILL) in Grenoble / France.
The result was:
\begin{equation} 
\label{eqn:Baumann}
q_{\rm n} = (-0.4\pm 1.1)\times 10^{-21}\,q_{\rm e},\quad (68\,\%\,\rm c.l.).
\end{equation} 
About the same time, another experiment was performed with ultra-cold neutrons (UCN) at the VVR-M reactor 
in  Gatchina / Russia \cite{Borisov1988}. The experiment had a discovery potential 
of $\delta q_{\rm n}=3.6\times 10^{-20}\,q_{\rm e}/\rm day$. Due to a measuring time 
of only three days and a certain misalignment of the device, the final result was:
\begin{equation} 
\label{eqn:Borisov}
q_{\rm n}=(4.3\pm 7.1)\times 10^{-20}\,q_{\rm e},\quad (68\,\%\,\rm c.l.). 
\end{equation} 
These direct measurements have in common that a beam deflection
$\Delta x$ of a charged particle 
depends on the square of its traveling time in an electric field $E$:  
\begin{equation}
\Delta x \propto\,\frac{q\,E}{m}\,{t^2},
\end{equation}
where $q$ is the charge and $m$ is the mass of the particle.

The measurement is carried out on the steep slope of the beam profile where the gradient
$\frac{dN}{dx}$ is maximum. At this point the beam deflection $\Delta x$ is most sensitive to a difference in the count-rate  
$\Delta N=(N_+-N_-)$ with respect to a reversal of the electric field from $E_+$ to $E_-$, and:
\begin{equation}
\label{eqn:Deflection}
\Delta x = \Delta N/\left ( \frac{dN}{dx}\right ).
\end{equation}
A neutron electric charge 
would cause a change in the count-rate according to:
\begin{equation}
q_{\rm n} \propto \frac{1}{E\,{t^2}} \left[ \Delta N/\left ( \frac{dN}{dx}\right )\right ].
\end{equation}

If $\Delta N$ is determined with a statistical uncertainty $\sqrt{2\,N}$
(assuming $\Delta N \approx 0$)
in a measuring time $\tau$, the sensitivity on $q_{\rm n}$ (i.e.\,the discovery potential) is:
\begin{equation}
\delta q_{\rm n} \propto \frac{1}{E\,\bar{t^2}} \left[ \sqrt{\frac{2\,N}{\tau}}/\left ( \frac{dN}{dx}\right )\right ].
\end{equation}
$\bar{t^2}$ now is the mean square transit time of a neutron velocity spectrum through the experiment. 
In first approximation, the obtained beam profile has a Gaussian shape. At the steep point of the slope:
\begin{equation}
\label{eqn:SteepSlope}
\frac{dN}{dx}\propto \frac{N}{w_0},
\end{equation}
where $w_0$ is the width of the beam profile. 
Therefore: 
\begin{equation}
\label{eqn:Sensitivity}
\delta q_{\rm n} \propto \frac{w_0}{E\,\bar{t^2}\sqrt{N}}.
\end{equation}

We consider (\ref{eqn:Sensitivity}) a figure of merit and compare the experiments with cold neutrons \cite{PhysRevD.37.3107}
and with UCN \cite{Borisov1988} in tab.\,\ref{tab:Comparison}.
Note that despite the high gain factors for $N$ and $w_0$ in the 
cold-neutron-experiment, the UCN experiment takes advantage of the much larger flight time of the ultra-cold
neutrons in the region of the electric field. Nevertheless, the sensitivity 
of \cite{Borisov1988} is 16 times smaller compared to \cite{PhysRevD.37.3107}. 

\begin{table}[h]
\begin{center}
\begin{footnotesize}
\begin{tabular}{|c || c | c | c | c | }
\hline
Experiment &$N\,[1/\rm s]$ & $w_0\,[\mu\rm m]$ & $E\,[\rm V/m]$ & $\bar{t^2}\,[\rm s^2]$\\
\hline	   
\cite{PhysRevD.37.3107}, CN	&	$3\times 10^4$	&   $30$ & $6\times 10^6$ &  $1.4\times 10^{-3}$\\
\hline
\cite{Borisov1988}, UCN		&	$3\times10^2$	& $700$ & $10^6$ &  $0.12$\\
\hline
\hline
Gain factor  &$\sqrt{100}$ & 23  & 6 & $1.16\times 10^{-2}$\\				
\hline
\end{tabular}
\end{footnotesize}
\caption{Comparison of the two recent experiments on $q_{\rm n}$. CN - cold neutrons, UCN - ultra-cold neutrons.
$\bar{t^2}$ for \cite{PhysRevD.37.3107} was estimated from $l=9\,\rm m$ and $\bar{v}=240\,\rm m/s$ as stated 
by the authors. The overall gain factor of \cite{PhysRevD.37.3107} is 16.}
\label{tab:Comparison}
\end{center}
\end{table}

The availability of highly intensive UCN sources (see \cite{ILL08Proceedings} and references therein for a comprehensive
overview) at different research 
centers such as PSI (Switzerland), ILL (France), LANL (USA), KEK (Japan), FRM\,II, or TRIGA Mainz 
(both Germany) will significantly improve the measurement sensitivity of many experiments and will also
trigger new experimental approaches. Besides our attempt we mention as one example \,the GRANIT, a follow-up project based on an UCN 
spectrometer of ultra-high resolution, which provides accurate studies of/with the quantum states of neutrons in Earth's 
gravity field \cite{Nesvizhevsky2006}. A test of \q is also proposed in that paper.

\section{Experimental setup}
Our experiment was installed at the TES beamline of the UCN-facility PF2 \cite{ILL} 
at ILL. UCN are guided via a 2.5\,m long 
stainless steel tube of 70\,mm diameter from the exit of the UCN turbine to the experiment.
To protect the turbine vacuum and for safety reasons, a $100\,\mu \rm m$ aluminum foil is permanently installed directly after
the exit of the turbine.

\begin{figure*}[t]
\centering{
\includegraphics[width=.5\textwidth, angle=0]{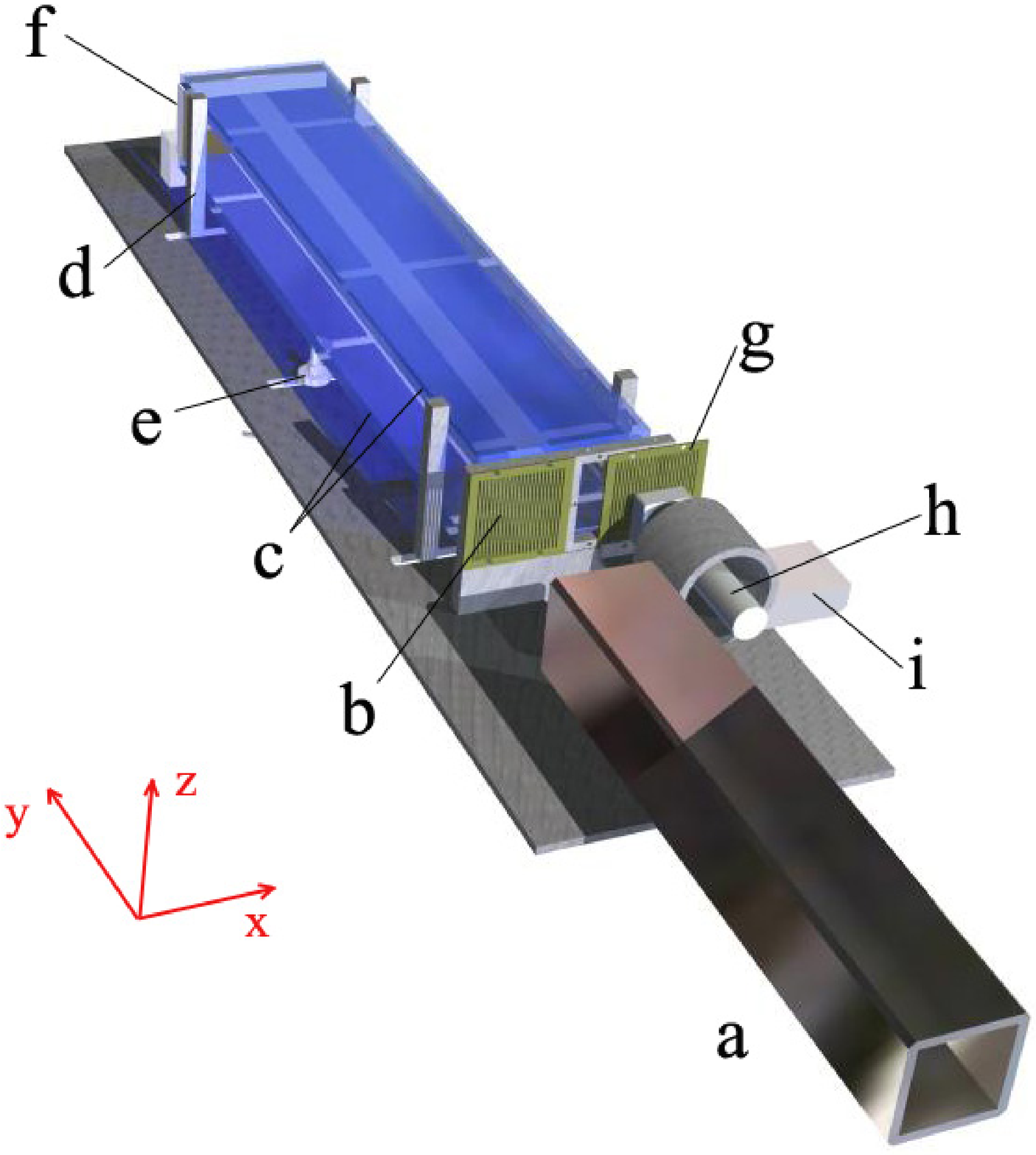}
\caption{\label{fig:Experiment} Draft of the experiment. a - rectangular guide, b - entrance grating, c - guide plates,
d - bars with noses, e - linear actuator, f - curved mirror, g - exit grating, h - detector, i - linear stage.}
}
\end{figure*}

Figure \ref{fig:Experiment} shows a scheme of the experiment; the $x,y,z$ directions are indicated;
$|z|$ is the direction of gravity. The vacuum is about $10^{-4}\,\rm mbar$ during all measurements. 
Information on the used materials is given in the following section.

\begin{figure*}[t]
\centering{
\includegraphics[width=.75\textwidth, angle=0]{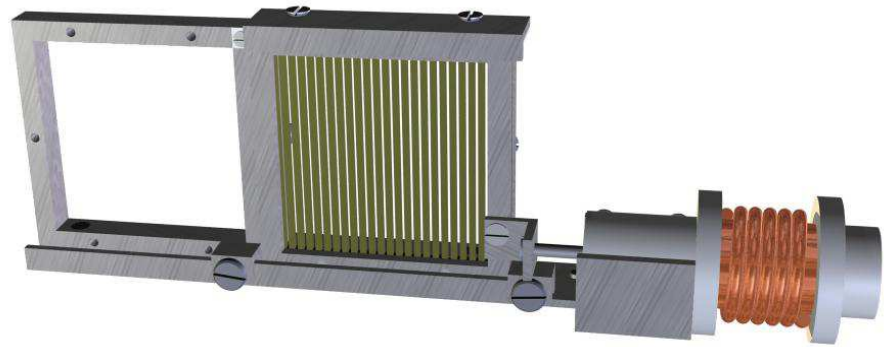}
\caption{\label{fig:Chopper}Draft of the exit (left) and the entrance (right) channel and the entrance grating (the exit grating is not shown here). For the ToF-measurements, a second entrance grating installed over the first one
is periodically moved by an actuator coil.}
}
\end{figure*}

At the entrance of the experimental chamber the circular beam cross section is converted 
into a rectangular shape by means of a nickel  guide (a) of inner dimensions $50\times50\,\rm mm^2$ and 
34\,cm length.
After the rectangular guide, the UCN pass an entrance grating (b), which is shown in detail also 
in fig.\,\ref{fig:Chopper}. The grating has an overall dimension 
of $50\times 50\,\rm mm^2$. The slits are 0.4\,mm or 0.7\,mm wide and
are separated by a distance of 1.5\,mm. The lattice constant is therefore 
1.9\,mm or 2.2\,mm. 
This entrance grating fragments the beam profile, and transmitted UCN are
reflected between two horizontal neutron guide plates (c).
The vertical distance between the plates is 50\,mm, the dimensions of the plates 
are: Length 500\,mm, width 100\,mm, thickness 8\,mm.

The upper plate lies on four adjustable nose-pieces sticking out of aluminum bars (d).
The lower plate is supported at three positions; on two nose-pieces and on
one linear actuator (e, see also fig.\,\ref{fig:Tilting1}) which is movable in $z$-direction. 

UCN of too high divergence in $x$-direction leave the device to the sides and are absorbed
by borated rubber in the experimental chamber.

The device can be described as an optical camera for ultra-cold neutrons:
A certain fraction of UCN reaches the end of the guide plates and encounter a mirror of cylindrical
shape (f) in the horizontal $x$-direction. The mirror has a
curvature of 500\,mm with dimensions $100\times 50\,\rm mm^2$ in $x, z$-direction.
It reflects the UCN and focuses the image of the entrance grating back onto
an exit grating (g). The latter is of same dimensions as the entrance grating. 
Together with a detector (h) mounted directly behind, it can be shifted along the $x$-direction
by a linear stage (i) in order to scan the reflected beam image.  
In the following, we refer to the detector together with the exit grating as the 'exit channel'.

The electrodes would be situated perpendicular to the guide plates generating an electric
field in the $x$-direction along the neutrons' flight path. In the present setup we did not implement
a HV system.

\section{Materials and alignment}
\label{sec:Materials}
{\it General remarks}:
We recall that the interaction of slow neutrons with matter can be described by a potential of the form: 
\begin{equation}
\label{eqn:Potential}
U=V-\imath W, 
\end{equation}
where $V$ and $W$ are derived from parameters of the material surface at 
which the neutrons interact via the strong force (see e.g. \cite{Golub}).
In general, $V$ is called the 'wall potential'.
It defines a critical velocity $v_c=\sqrt{{2\,V}/{m_{\rm n}}}$ below which a neutron
is reflected under any angle of incidence from the surface. The non-vanishing probability of absorption is described by the imaginary part
of (\ref{eqn:Potential}).
As an example we cite the values for nickel:  $U=(252 - \imath\,0.032)\,\rm neV$ and $v_c=7\,\rm m/s$ \cite{Golub}.\\[5pt]
{\it Gratings}: For the gratings, a material of high UCN absorption probability is required, which can also
be machined with high precision. We chose titanium of 1\,mm thickness for the first. Ti is an effective UCN absorber
with a negative wall potential of $V=-51\,\rm neV$ and an absorption cross section
of 6.9\,barn at $v_{\rm n}=2200\,\rm m/s$ \cite{Golub}. 
The transmission probability for 1\,mm Ti is $10^{-5}$ for UCN of v=7\,m/s. 
The gratings were manufactured by CNC wire-erosion with a precision
of $10\,\mu \rm m$.

{\it Guide plates}: High specularity is required for the neutron guide plates, because diffusive reflection on the surface
will lead to a misalignment of the UCN trajectories. We have chosen 
float-glass for the guides. Off-the-shelf, this material provides
a high quality for surface roughness, typically on the order of $\delta\sim\,\rm 10\,\AA$. 
Note that the wavelength of UCN is on the order of some $\lambda \sim 100\,\AA$. In first order 
the probability for non-specular
UCN reflection scales with $(\delta/\lambda)^2$ \cite{Steyerl1986}. Therefore 
float-glass is of first choice\footnote{Note that high-quality UCN guides with an even lower surface roughness about
1\,\AA\, have been used in \cite{Plonka:2007kt,Nesvizhevsky2007}.}. 
Its wall potential ($V=95\,\rm neV$) can be increased by coating of the reflecting
surface with a non-conductive material\footnote{Nickel for the coating as well as titanium for the gratings might not be suitable materials 
once the high voltage is implemented.} like BeO ($V=242\,\rm neV$, used 
e.g.\,in \cite{Borisov1988}), 
diamond-like carbon ($V=250\,\rm neV$, \cite{Atchison:2005kl}) or cubic boron nitride ($V=305\,\rm neV$, \cite{cbn2009}).

{\it Mirror}: A material of high wall potential has to be used for the surface of the mirror, because 
UCN with their main velocity component $v_y$ have to be reflected. We chose
N-BK 7 glass, which was coated with a $2000\,\rm\AA$ layer of nickel.
The surface finishing of the glass is 10\,\AA\,and the precision of the curvature radius is
$\delta R/R=50\,\mu \rm m/500\,\rm mm=10^{-4}$.

{\it Detector}: The detector is a lithium-doped
glass scintillator (GS\,10 see \cite{GS10}) of dimension $30\times 30\,\rm mm^2$ which is coupled via a transition light guide
to a photomultiplier. It does not cover 
the whole area of the exit channel\footnote{The detector was on loan from LPC Caen / France.}. Therefore 
we placed the scintillator {\it centered} behind the exit grating with a shielding of borated rubber around it.
With respect to its chemical composition, the scintillator has a wall potential comparable to float-glass. 
Therefore only UCN with velocities higher than 4.3\,m/s can be detected.

{\it Alignment}: The adjustment of the entrance grating and the mirror
with respect to the TES beamline was done with a laser.
The guide plates were adjusted parallel to each other by a high-precision water-level. 
After closing the experimental chamber, a 
fine-tuning was performed by means of the linear actuator and with 
the observed neutron signal (see sec.\,\ref{sec:Modulation}).

\begin{figure*}[t]
\centering{
\includegraphics[width=.5\textwidth, angle=270]{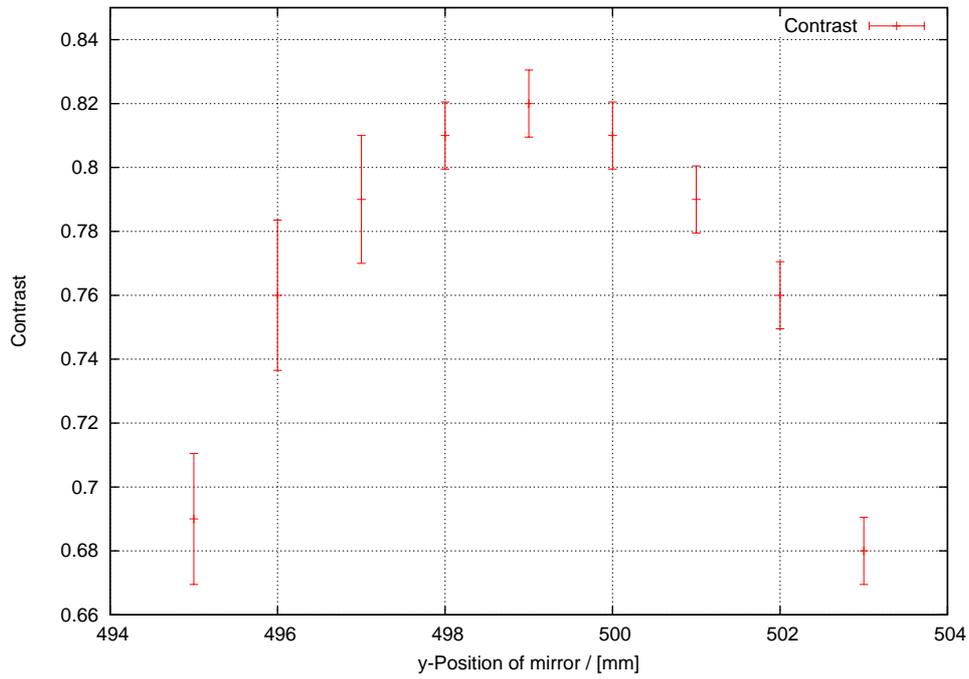}
\caption{\label{fig:LightModulation}Measured contrast of the light-modulation over the $y$-position of the mirror. The mirror is placed 
at the position of the maximum contrast at $y=499\,\rm mm$.}
}
\end{figure*}

The mirror was placed about its focus point $\sim$ 500\,mm behind
the entrance grating. The optimum position was found with a light-optical method before closing the experimental chamber:
We placed an array of LEDs before the rectangular guide and a photodiode at the position of the detector.
By moving the exit channel with the linear stage, we observed a modulation of the light intensity.
The average contrast $\frac{{I}_{\rm max}-{I}_{\rm min}}{{I}_{\rm max}+{I}_{\rm min}}$ of this curve is shown in fig.\,\ref{fig:LightModulation}.
The mirror was placed at the $y$-position of the maximum contrast. 
\section{Modulation measurements}
\label{sec:Modulation}
\begin{figure*}[t]
\centering{
\includegraphics[width=.5\textwidth, angle=270]{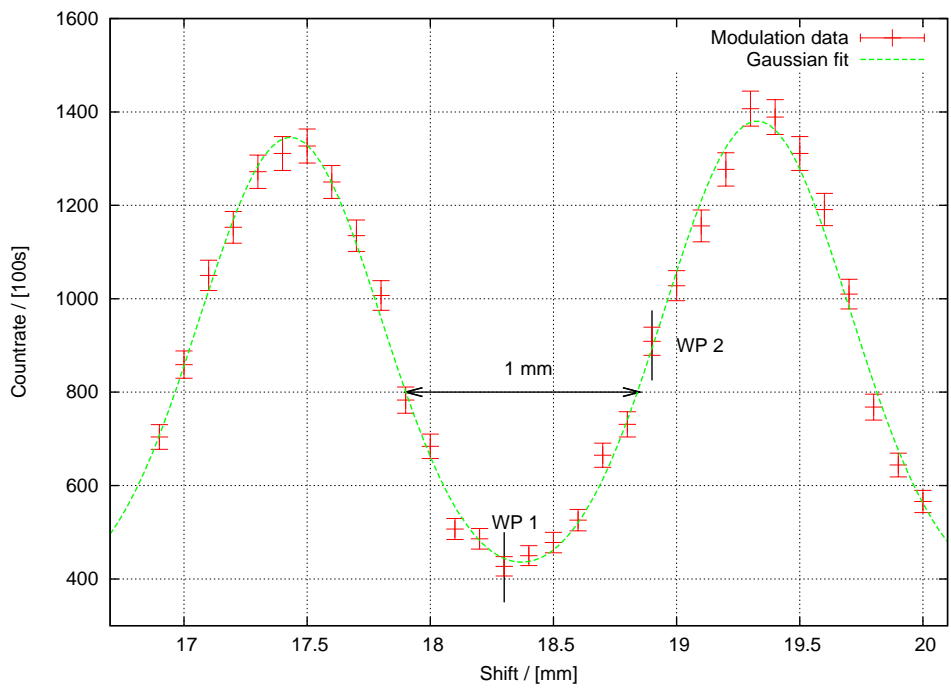}
\caption{\label{fig:Modulation}Measurement of the UCN modulation with the gratings of slite size 0.4\,mm and 
the Gaussian fit to the data. Also indicated are the two working points WP\,1 and WP\,2.
The indicated distance corresponds to the shift $2\,\delta x$, at which the count-rate approximately doubles
with respect to the value at WP\,1; see also text.}
}
\end{figure*}

Figure \ref{fig:Modulation} shows the UCN count-rate 
with the gratings of slit size 0.4\,mm. Starting from the initial position 
at $x=20\,\rm mm$ the 
exit channel is moved in steps of 0.2\,mm away from the entrance grating and
each position is measured for 100\,s. We observe a modulation of the count-rate.
The data are fitted by two Gaussian functions with the offset $N_0$:
\begin{equation}
{N}(x)=N_0+\sum_{i=1}^{2} \frac{N_i}{w_i \sqrt{2\,\pi}}\,e^{-\frac{(x-x_i)^2}{2\,w_i^2}}. 
\end{equation}
The fit parameters are summarized in tab.\,\ref{tab:Fit}, $\chi^2_{\rm red.}=1.3$.
\begin{table}[h]
\begin{center}
\begin{tabular}{| l || c | c | c |}
\hline
&$N_i$ [1/100\,s]&$w_i$ [mm]&$x_i$ [mm]\\
\hline	
$i=1$ &	964 (60)	&0.375\,(15)	&19.327\,(7)\\	
$i=2$ &	917 (60)	&0.369\,(16)	&17.432\,(8)\\
\hline	   
\end{tabular}
\caption{The fit parameters of the two Gaussian functions. The offset is 
$N_0=356\,(32)/\rm100\,s$.}
\label{tab:Fit}
\end{center}
\end{table}
The distance between the two peak positions, 1.89\,(1)\,mm, corresponds to the lattice constant of the
grating: (0.4+1.5)\,mm. From the minimum and the maximum count-rate (at $x=18.38\,\rm mm$ and $x_1$) we derive 
the contrast $C=\frac{{N}_{\rm max}-{N}_{\rm min}}{{N}_{\rm max}+{N}_{\rm min}}=0.53\,(1)$.
This maximum value was found after a fine-tuning of the lower guide plate 
(i.e.\,by a tilt of $<0.1\,\rm mm$ in $|z|$-direction) by means of the linear actuator.

The count-rate at the steep slope of the modulation (we look at $x=18.95\,\rm mm$) is ${N}=9.7\,(1)/\rm s$,
and the gradient at this point is $\frac{d N}{d x}=16.5\,(5)/\rm mm\,s$. The average width of the obtained beam 
profile is $2\,\bar{w_i} \simeq 0.74\,\rm mm$.
 
\section{Tilt measurements}
\label{sec:TiltMeasurements}
\begin{figure*}[t]
\centering{
\includegraphics[width=.5\textwidth, angle=270]{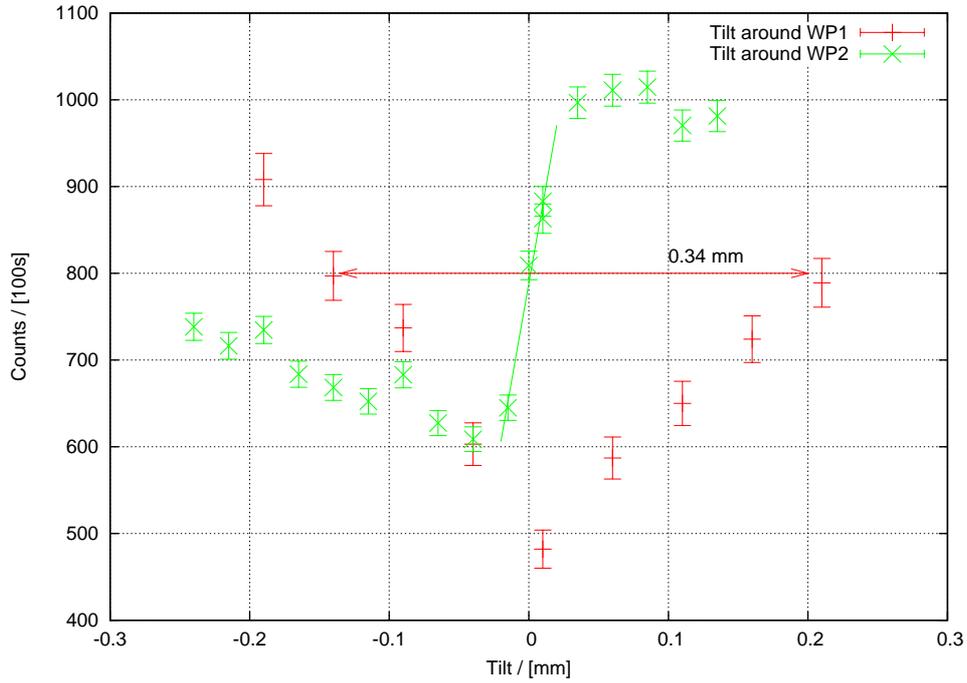}
\caption{\label{fig:Tilting}The count-rates at the working points WP\,1 and WP\,2 when the lower guide plate is tilted.
The indicated distance corresponds to the tilt $2\,\delta h$, at which the count-rate approximately doubles
with respect to the value at $\delta h=0$. The linear fit to the data for tilting around WP\,2 is indicated; see also text.}
}
\end{figure*}

We define two working points (WP) to investigate the influence of a tilt of the
lower guide plate on the observed count-rate. The working points are indicated in fig.\,\ref{fig:Experiment}. 
We chose $\rm WP\,1=18.3\,mm$ at the minimum and $\rm WP\,2=18.9\,mm$ at the steep slope of the 
modulation\footnote{These points do not fully coincide with the ones obtained from the fit: During the experiment,
only a preliminary analysis of the modulation curve was carried out.}. 
Figure \ref{fig:Tilting} shows the observed count-rates at these working points when the lower guide plate is tilted 
in $\pm\,z$-direction about $\delta h=0.2\,\rm mm$ in steps of 0.025\,mm. 
Around WP\,1, the count-rate increases in both directions from the minimum. 
At WP\,2 the count-rate rises or falls on a steep 
slope\footnote{We mention that a similar behavior was observed when the lower plate was tilted 
at the points of the maximum of the modulation or at its steep slope of opposite gradient 
(i.e.\,at $x=19.35\,\rm mm$ or at $x=19.75\,\rm mm$). In the first case, we observed a decreasing of 
the count-rate in both directions from the maximum and in the second case, the steep slope is just of opposite orientation.}.

\begin{figure*}[t]
\centering{
\includegraphics[width=.5\textwidth, angle=0]{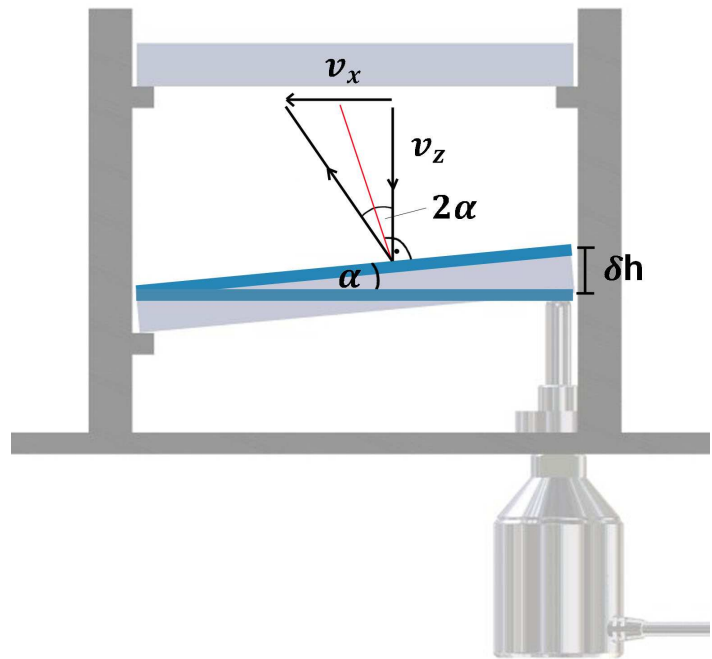}
\caption{\label{fig:Tilting1}Illustration of the relation between tilt and shift. 
The lower plate can be tilted in both $z$-directions by means of the linear actuator. 
Due to the tilt $\delta h$, a UCN with $v_z$ before reflection gains a velocity component $v_x$ after reflection.}
}
\end{figure*}

This can be understood in the following way: Undergoing a reflection at the tilted lower guide plate,
a neutron gains velocity in the $x$-direction as illustrated in fig.\,\ref{fig:Tilting1}. 
The resulting $v_x$-component depends on the tilt $\delta h$, on the $z$-velocity and on 
the number of reflections of the neutron.

At WP\,1 and without tilting, most of the neutrons hit the grating material of the exit channel and therefore a minimum
in the modulation curve is observed. With a shift of the neutron trajectories caused by a tilt of the lower guide plate,
the count-rate is increased in both tilting directions. In other words: A tilt of the guide plate is in some sense equal to
a shift of the exit channel. In this way, also the data around WP\,2 can be interpreted: Due to the tilt of the lower plate,
the neutron trajectories are shifted towards the minimum or the maximum position adjacent to WP\,2. 

A simplified model is used to further understand the results: 
We assume that neutrons with velocity components $v_y, v_z$ 
undergo only one reflection on the lower plate after about half of their total flight time $t$ between the entrance
and the exit grating. The lower plate is tilted by $\delta h$ and the resulting tilt angle is 
$\alpha<<1$ (see fig.\,\ref{fig:Tilting1}). 
For mono-energetic UCN, the induced shift of a neutron arriving at the exit channel is:
\begin{equation}
\label{eqn:ShiftTilt}
\delta x = 2\,\alpha\,v_z\, \frac{t}{2}, \quad \mbox{or}\quad
\delta x = \frac{\delta h}{b}\,\frac{{v}_z}{{v}_y}\,l.
\end{equation}
$l$ is the length of the guide plate (500\,mm) and $b$ its width (100\,mm). 
For non-monochromatic UCN, $\delta x$ will be different for each neutron.
We can expect that due to the resulting broadening of the beam profile the 
count-rate will not rise or fall to the maximum and minimum values adjacent to WP\,2 
(i.e. 14/s and 4/s) when the lower
plate is tilted further. This is indeed observed for tilt values $\delta h=(-0.2 ; +0.1)\,\rm mm$, where
a saturation of the count-rates (about 10/s and 7/s) shows up (see fig.\,\ref{fig:Tilting}).

We further test our assumption 'tilt equals shift' by putting numbers into the model: 
A tilt of $\delta h= 0.17\,\rm mm$ around WP\,1 (see fig.\,\ref{fig:Tilting}) increases the count-rate to $N=8/\rm s$. Approximately
the same enhancement is observed when the exit channel is shifted by $\delta x=0.5\,\rm mm$ (see fig.\,\ref{fig:Modulation}). 
We take $t=0.152\,\rm s$ (from the ToF measurements described in sec.\,\ref{sec:ToF}) and
$\alpha =1.7\times 10^{-3}\,\rm rad$. Further we assume a quadratic velocity dependence for 
$v_z$ between zero and 4.3\,m/s, the critical velocity of float-glass, resulting in a mean value of $\bar{v}_z=3\,\rm m/s$.
With these values and from the left part of (\ref{eqn:ShiftTilt}) we calculate $\delta x= 0.78\,\rm mm$. This is more or less in
agreement with the expected value of 0.5\,mm. 

One could argue that under the assumption of $\bar{v}_z=3\,\rm m/s$ more than one reflection on the lower guide plate occurs and 
that the resulting shift of a UCN due to the tilting should be larger. However, a UCN beam shows a forward peaking on transmission 
through guides depending on the specularity of reflection \cite{Golub}. This angular distribution $f(\theta_z)$ was not taken into account in our model, but
it will lead to smaller values for ${v}_z$. Therefore, the derived value $\delta h=0.78\,\rm mm$ has to be interpreted as an upper limit. 

We focus again on the working point WP\,2 and investigate the influence of a tilt upon the count-rate around this point.
A linear fit to the data as indicated in fig.\,\ref{fig:Tilting} ($\chi^2_{\rm red}=1.39$) 
yields the gradient $\frac{d N}{d h}=(91\pm9)/\rm mm\, s$.

What is the meaning of this value? Apart from forces acting on the neutrons perpendicular to their flight direction (which we aim
to detect with such a device), 
any change of the experimental environment e.g.\,by vibrations or electrostatic forces can also give rise to misaligned UCN trajectories,
 as the guide plates might be tilted by small amounts.This would lead to false effects.

In a measuring time $\tau$, a change in the UCN count-rate 
due to a tilt $\delta h$ has to be smaller than the statistical uncertainty on $\Delta N$. This is mandatory to verify a shift of UCN trajectories
which may arise from 
perpendicular forces. Therefore:
\begin{equation}
\label{eqn:SystematicTilt}
\delta h<\sqrt{\frac{{2\,N}}{\tau}}\times \left (\frac{d N}{d h} \right )^{-1}.
\end{equation}
Assuming a measuring time of 100\,s, $N\approx 10/\rm s$ and the gradient from above, it follows $\delta h<5\,\mu\rm m$.

Equation (\ref{eqn:SystematicTilt}) comprises a most important systematic effect. 
Note that $\frac{d N}{d h}\propto N$, as in (\ref{eqn:SteepSlope}). Therefore the acceptable tilt over a measuring time
will scale with $\delta h \propto {N}^{-1/2}$. 

So far we considered tilting as a systematic effect, which leads to misaligned UCN and therefore affects the count-rate. A tilt of the guide plates
could occur for example if an electric field is reversed and the attached guide plates are slightly shifted in $z$-direction due to the electrostatic forces between the electrodes. 
Note, that such a force is on the order of 0.2\,N, assuming an electric field of $10^6\,\rm V/m$ and dimensions as given in \cite{Borisov1988}. 
In fact, the authors of that paper observed such tilting effects and backtracked them to electrostatic forces.

Tilting of the {\it whole} device however can also be used to calibrate the experiment and to derive a value for $\bar{t^2}$, independent of 
the time-of-flight measurements presented in the next section. We will come back to this important point in the appendix A.

\section{Time-of-flight measurement}
\label{sec:ToF}
We now address to the time-of-flight (ToF) of the UCN through the device. A chopper system
was built with a second entrance grating (see fig.\,\ref{fig:Chopper}), that can be moved
periodically in $x$-direction across the first one. 
To gain in statistics, the 0.7\,mm gratings at the entrance, and no grating at the exit channel was installed.

\begin{figure*}[t]
\centering{
\includegraphics[width=.75\textwidth, angle=0]{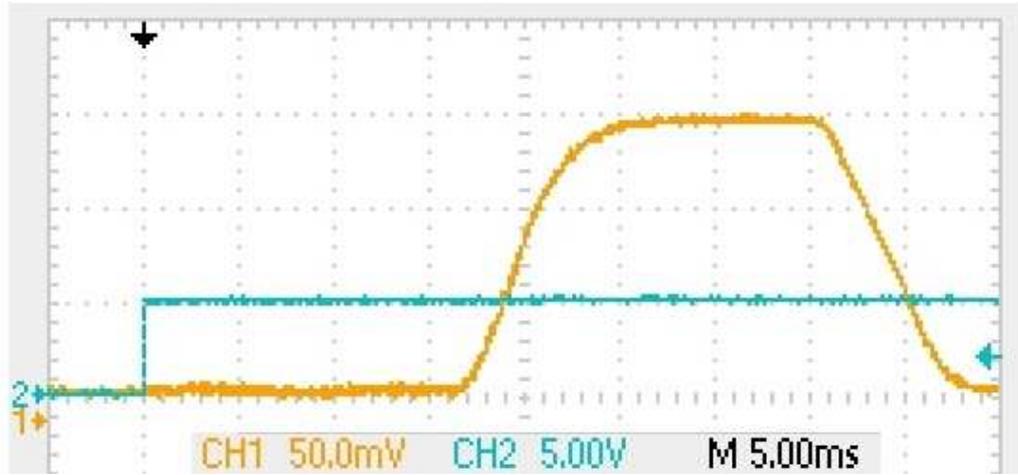}
\caption{\label{fig:Choppersignal} Opening function of the chopper (1, yellow) and the starting signal 
of the ToF measurement (2, green).}
}
\end{figure*}

In order to measure the opening function of the chopper, we placed an array
of LEDs in front of, and a photodiode behind the chopper. The opening function is 
shown in fig.\,\ref{fig:Choppersignal}. The chopper is fully open for about 15\,ms. Taken into account the slopes
we derive an effective opening
time of 18\,(2)\,ms. The starting signal for the ToF is delivered
from a frequency generator, which also drives the actuating coil via an amplifier. 
This signal (see fig.\,\ref{fig:Choppersignal}) starts the data acquisition 30\,(2)\,ms before the chopper is fully open. The
time-offset has to be considered in the analysis of the ToF data.

\begin{figure*}[t]
\centering{
\includegraphics[width=.5\textwidth, angle=270]{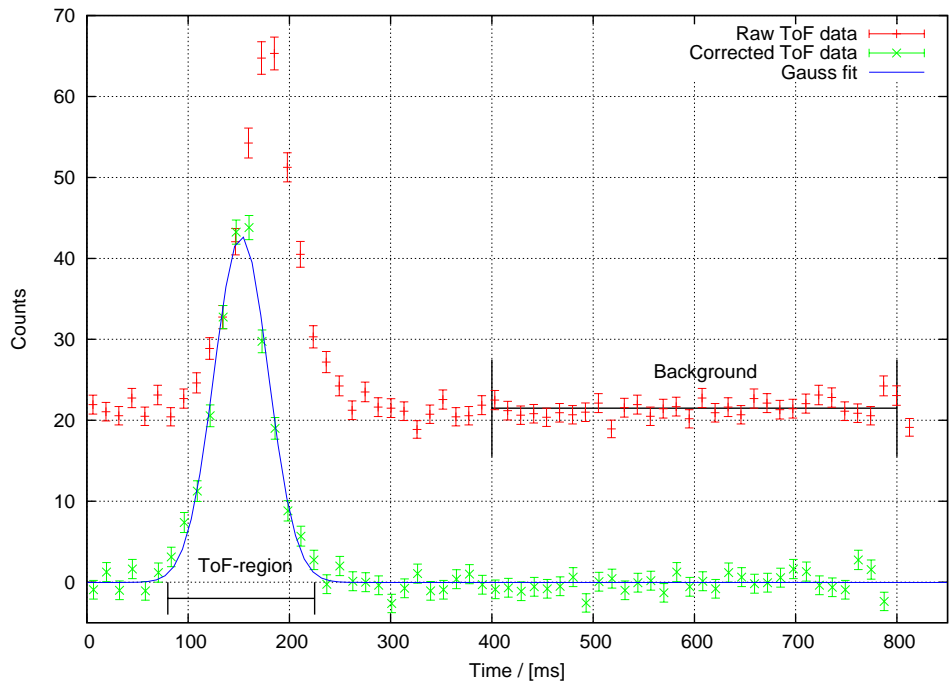}
\caption{\label{fig:ToF} Raw (red) and corrected (green) ToF data. The regions for the background fit and for the calculation
of $\bar{t^2}$ are indicated.}
}
\end{figure*}

Figure \ref{fig:ToF}  shows the raw ToF data (red) taken in one hour of measuring time. The duty cycle of the chopper
is $2\,\%$. First, the background is fitted in the indicated region
with a constant  $bg=(21.5\pm0.2);\,\chi^2_{\rm red}=0.84$. Second, $bg$
and the time-offset are subtracted from the raw spectrum. The corrected data (green) are also shown in fig.\,\ref{fig:ToF}.
A Gaussian function is fitted to the corrected 
data, resulting in $y_0=(-0.02\pm0.2)$, the net area $A=2871\,(67)$, the mean time-of-flight 
$\bar{t}=152.5\,(7)\,\rm ms$ and the width $w=26.8\,(7)\,\rm ms;\,\chi^2_{\rm red}=1.16$.
To obtain $\bar{t^2}$, we analyze the data in the $5\,\sigma$ region around $\bar{t}$ as indicated in fig.\,\ref{fig:ToF}:
\begin{equation}
\bar{t^2}=\frac{\sum_i{t_i^2\,C_i}}{\sum_i{C_i}} = 0.022\,(1)\,\rm s^2,
\end{equation}
where $C_i$ is the count-rate of the ToF bin $t_i$. 

A velocity spectrum is derived from the corrected data using the transformations: $t\rightarrow l/t$ and 
$\frac{dN}{dt} \rightarrow\frac{dN}{dv}\frac{dv}{dt}=\frac{dN}{dv}\frac{t^2}{l}$, see e.g.\,\cite{Altarev2008}. 
For the mean flight length we take $l=1100\,\rm mm$, which follows from $\bar{v}_z=3\,\rm m/s$ 
(see sec. \ref{sec:Modulation}) and $\bar{v}_y=6\,\rm m/s$ 
(assuming a quadratic velocity dependence between 3.2\,m/s, the aluminum cut-off of the safety window, 
and 7\,m/s, the nickel cut-off of the mirror). 
Note, that a detailed analysis would have
to take into account the opening function of the chopper as well as a more realistic assumption on the
velocity distribution. However, we are interested in a qualitative understanding of the data for which this 
basic approach is sufficient.

\begin{figure*}[t]
\centering{
\includegraphics[width=.5\textwidth, angle=270]{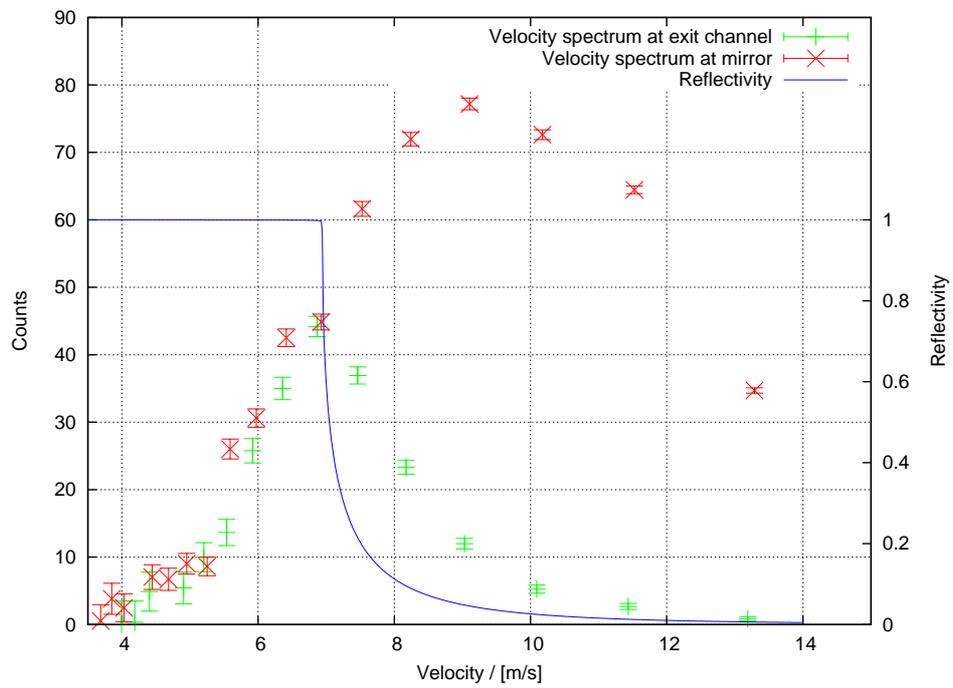}
\caption{\label{fig: VelocitySpectra}Velocity spectrum, derived from the ToF data. Green: Spectrum measured at the
exit channel; Red: Spectrum measured at the position of the reflecting mirror. Blue: Reflectivity 
$|R^2|$ of nickel versus the UCN velocity.}
}
\end{figure*}

The obtained velocity spectrum (green) is shown in fig.\,\ref{fig: VelocitySpectra}.
We state: a) The minimum detected velocity is about $v=4\,\rm m/s$. This has two reasons: First, the aluminum safety window at the exit of the turbine only transmits neutrons with $v>3.2\,\rm m/s$. Second, the minimum velocity of detectable neutrons is $4.3\,\rm m/s$ due to the wall potential of the scintillator. 
b) The maximum about $v=6.9\,\rm m/s$ corresponds to the critical velocity of the nickel coated mirror.
c) There is a significant contribution from neutrons above the nickel cut-off. 

For two reasons these neutrons behind the Ni cut-off show up 
in the spectrum: 1) At the TES beamline, also neutrons with velocities higher than 7\,m/s are delivered from the 
UCN turbine. This is shown in fig.\,\ref{fig: VelocitySpectra} (red). In this measurement the detector was 
placed at the position of the mirror. Neutrons with velocities up to 15\,m/s, so called very-cold neutrons (VCN), contribute to the
spectrum. 2) The reflectivity for a UCN of energy $E$, impinging perpendicular on a surface
with a wall potential $U=V-\imath W$ is given by \cite{Golub}:
\begin{equation}
\label{eqn:Reflectivity}
|R|^2= \left |\frac{\sqrt{E}-\sqrt{E-U}}{\sqrt{E}+\sqrt{E-U}}\right |^2.
\end{equation}
The reflectivity curve of the nickel mirror surface is also shown in fig.\,\ref{fig: VelocitySpectra}. Note, that due to the non-step-wise character
of $|R|^2$, also UCN above the Ni cut-off can be reflected.
Apart from losses of the back-reflected UCN and the different dimensions of the mirror and the detector surfaces, 
the obtained velocity spectrum at the exit channel is in principle the convolution of the reflectivity curve with the 
velocity spectrum at the position of the mirror. 
\begin{table*}[t]
\begin{center}
\begin{footnotesize}
\begin{tabular}{| l || c | c | c | c || c | c |  c | c  | c  | c  |}
\hline
&\multicolumn{4}{| c ||}{Parameters and dimensions}& \multicolumn{6}{| c |}{Measured values}\\
\hline
&UCN flux&Flight&Slit	&Detector& Contrast &Countrate&Gradient&$\bar{t^2} \,[\rm s^2]$&$\bar{v} \,[\rm m/s]$&Offset\\
 
&$[\rm cm^{-2} s^{-1}] $&path [\rm m]&size [\rm mm]&area / [cm$^2$] &&$N$  [1/s]&$\frac{dN}{dx} \,[\rm 1/mm \,s]$&&&$N_0$  [1/s]\\
\hline
Our experiment	      & $8\times 10^2$&1	& 0.4& 9 & 0.53 & 9.7  & 16.5  & 0.022 & 7&3.6\\
\hline
\cite{Borisov1988} & $6\times 10^3$&1.8	& 0.7&20& 0.28 & 280 & 205& 0.12 & 5&$\approx 100$\\
\hline 
\hline 
Comparison I    & 8 & 0.56 & 1.75 & 2.2 &0.5 &29&12.5&5.5&&28\\
\hline
\hline
Comparison II   & \multicolumn{4}{| c ||}{$17.25\times 1.5=26$}& \multicolumn{6}{| c |}{}\\
\hline
\end{tabular}
\caption{Comparison of the present and the former UCN experiment \cite{Borisov1988}.
The count-rate and the gradient for \cite{Borisov1988} are averaged over the two exit
channels of that experiment.
The value 1.5 in 'Comparison II' is an additional geometrical factor which arises from the transition of the circular to the rectangular guide 
cross section in our experiment; see also text.}
\label{tab:Comparison2}
\end{footnotesize}
\end{center}
\end{table*}

\section{Analysis}
We analyze the experiment and compare our results in tab.\,\ref{tab:Comparison2} with the former UCN experiment \cite{Borisov1988}.

{\it Contrast}: The ratio of the contrast values of the two experiments is $0.5$. To explain this difference we assume, that the contrast is mainly affected
by the roughness of the guide plates: Roughness leads to diffusive reflection (see e.g.\,\cite{Heule2008} for an overview and experimental results), 
to a misalignment of UCN trajectories and to a broadening of the obtained image. Let us consider further that the roughness 
of the plates in both experiments is about the same. We cannot really prove this assumption at the moment but we state that a diffusive reflection factor 
of 1.4\,\% as given in \cite{Borisov1988} is quite a realistic value for float-glass in UCN applications (see again \cite{Heule2008}).
How much a UCN is misaligned in $x$-direction on its way through the device scales with the number of reflections on the 
surface and therefore with the flight path.
The ratio between the flight paths of the two experiments, 0.56, corresponds to the ratio of the contrast values. In this view, roughness
of the surface has to be considered as an important parameter.

The quantum mechanical broadening in $x$-direction due to Heisenberg's uncertainty principle is:
\begin{equation}
\label{eqn:Heisenberg}
\Delta \chi_{\rm q.m.} \approx \frac{h}{m_{\rm n}}\,\frac{t}{w_0}.
\end{equation}
This value is about 0.15\,mm in our setup (for the grating with $w_0=0.4\,\rm mm$). Therefore quantum mechanical broadening is 
not negligible and a further decrease of
the slit size will not consequently result in a higher contrast or a steeper gradient.

{\it Count-rate}: The ratio of the count-rates in the two experiments is 29. We have to take
into account several factors to understand this value: 
The UCN flux at the former VVR-M reactor was about a factor 5 smaller than at the most intensive beamline of PF2 (which is
called EDM):
\begin{eqnarray}
\label{eqn:FluxComparison}
6\times 10^3\,\rm cm^{-2} s^{-1}  &{\rm for}\,\,v<7.8\,\rm m/s  &\mbox{VVR-M\,}\nonumber\\ 
3.3\times 10^4\,\rm cm^{-2} s^{-1}&{\rm for}\,\,v<7.0\,\rm m/s  &\mbox{PF2, EDM}\,\nonumber 
\end{eqnarray}
Further, at ILL, the UCN flux at TES is about a factor 40 smaller compared to EDM\footnote{We determined this ratio by a comparison of UCN transmission measurements through a nickel foil at both beamlines. Independently, this ratio was found 
in ToF measurements by M. Daum et al. \cite{Daum2008} at these two beamlines.}, hence about  $8\times 10^2\,\rm cm^{-2} s^{-1}$.

In first order, the observed count-rate rate is proportional to the solid angle covered
by the mirror and the detector area over the flight path. A larger slit size will increase the 
count-rate\footnote{Experimental proof of these two statements during our measurements: 
With a mirror of only half dimension in $x$-direction (i.e.\,5\,cm), the detected 
count-rate also decreased by a factor of 2. 
The modulation was also measured with the gratings of slit size
0.7\,mm (before: 0.4\,mm). In this configuration we observed an increase in count-rate about a factor of 2.}. 

From these arguments, the comparison
factors I in tab.\,\ref{tab:Comparison2} with respect to flight path, slit size and detector area  
are derived. Together with an additional loss factor of 1.5, arising from the transition from the circular
to the rectangular guide ($7\,\rm cm\, \varnothing  \rightarrow 5\,\rm cm\, \Box $), the expected loss factor for our experiment is about 26 (Comparison II in tab.\,\ref{tab:Comparison2}).
This is quite comparable to the observed ratio of the count-rates as stated above.

{\it Gradient}: With respect to (\ref{eqn:SteepSlope}), 
we would expect a ratio of the gradient values comparable to the ratio of the count-rates, as the width $w_0$ of the obtained image 
for both experiments is about the same (0.74\,mm in our experiment, see tab.\,\ref{tab:Fit}, and about 0.8\,mm in the former setup \cite{Borisov1988}). Note however,
that the value $dN/dx=205/\rm mm\,s$ in the former experiment is the one used for the analysis. 
Apart from a few runs with higher gradients, this cited number is the average value over the full beam time, where  $dN/dx$ was continuously 
monitored \cite{Plamen2009}. The determined ratio of 12.5 has to be interpreted
therefore as a lower limit.

{\it Time-of-Flight:}
There is a difference in the $\bar{t^2}$ value of both experiments by a factor of 5.5. This arises from a larger
flight length and a smaller mean UCN velocity in the former setup. We were limited to detect UCN only above the wall potential of the scintillation
detector; 4.3\,m/s. This could be overcome by using another detector material like $^{10}\rm Boron$ with a wall potential 
close to zero in combination with a semiconductor (see \cite{Jakubek:2009df,Jakubek:2009cv}). 
To overcome the velocity limitation arising from the safety window (in our case: aluminum with $V=54\,\rm neV$),
the experiment would have to be shifted up in height by about half a meter in order to slow down the UCN. 
We state that the sensitivity increases significantly, if the mean velocity $v_y$ is about 2--3\,m/s.

{\it Offset and background:}
The ratio of the offset values, 28, corresponds to the ratio of the count-rates of the two experiments. This is another evidence for our above made
assumption, that the contrast is mainly affected by the roughness of the guide plates. In other words: A higher count-rate will lead to an increase
of misaligned UCN, which contribute to $N_0$ in the modulation curve. Therefore this offset cannot be interpreted as a pure background but as an 
immanent characteristic of the device, which mainly depends on the surface quality of the guide plates. The background itself was
measured independently (by placing absorbing materials between the guide plates) and was found to be negligible in our experiment.

\section{Summary}
We revisited the experiment of \cite{Borisov1988}, a test of the electric charge of the neutron, 
at the beam position TES of PF2 at the ILL, and we reproduced the relevant 
parameters of the former setup in good agreement. We investigated the apparently most important systematic 
effect\footnote{Other systematic effects are discussed 
in \cite{PhysRevD.37.3107} and \cite{PhysRevD.25.2887} in consideration of using cold
neutron. The conclusions are conferrable to UCN.} 
of a tilt of the guiding components. The acceptable limit for such a tilt would be on the $\mu\rm m$-level for our setup and scales with $N^{-1/2}$. 

We used a light-optical method to adjust the reflecting mirror to its optimum position. As an 
online-monitor for the alignment of the device, independent of the observed UCN count-rate, this 
method will be further developed.

A direct measurement of the mean transit time and $\bar{t^2}$ between the entrance grating and the exit channel, i.e.\,in the relevant
region along the situated electrodes, was demonstrated by means of a dedicated chopper system. 

In view of a test of the electric charge of the neutron we state, that 
the expected gain in UCN density and flux at one of the new UCN sources like 
\cite{PSI_UCN} is about a factor 100 compared to PF2--EDM at ILL. With an upgraded experiment following
the approach of \cite{Borisov1988} and with control, monitoring and active compensation of the tilt effect, 
a sensitivity on \q on the $10^{-22}\,q_{\rm e}$ level within a measuring time of a couple of weeks can be reached.



\appendix

\section{Calibration by gravitational shift}
\label{sec:Grav_Shift}
If the {\it whole} device in fig.\,\ref{fig:Experiment} (i.e.\,gratings, guides plates and attached electrodes, mirror and detector) is tilted by an angle $\alpha$ with respect to the $x$-axes, 
the neutron image is shifted by gravity along the direction of the electric field. 
This effect was demonstrated in \cite{Borisov1988}
and allows an independent determination of the mean square transit time.
Assuming monochromatic UCN, there is a relation between the tilt $\alpha$ and the shift $\Delta x$:
\begin{equation}
\label{eqn:Calibration2}
\Delta x = \frac{g\,{t^2}\, \sin \alpha}{4}.
\end{equation}
As a result, the count-rate will change according to:
\begin{equation}
\label{eqn:Calibration3}
\Delta N = \frac{g\,{t^2}\, \sin \alpha}{4}\frac{d\,N}{d\,x}.
\end{equation}
The authors of \cite{Borisov1988} carried out such a calibration under
the assumption of a non-monochromatic velocity distribution, with the geometrical parameters of their device and a diffusive reflection factor 
of 1.4\% for the guide plates. From the obtained calibration curve and the leveling of their device they 
derived $\bar{t^2}=0.12\,\rm s^2$. This value was in good agreement with the result obtained from independent time-of-flight measurements\footnote{In
\cite{Borisov1988} a value of $\bar{t^2}=0.3\,\rm s^2$ is reported. Note, that this is the mean square transit time of
the UCN through the {\it whole} device obtained from velocity spectra which were measured between the entrance and at the exit of the device. A mean velocity
5\,m/s was found. From this it follows a mean square transit time of about $0.12\,\rm s^2$ along the electrodes; $l=2\times 0.9\,\rm m$.}.

\section{Gravitational attraction}
Following a remark of A.\,Franck\,\cite{Frank2009}, by placing massive bodies along the UCN flight path (i.e.\,at the position of the electrodes), the interaction of neutrons with gravity could be measured with a similar setup as described above. 
Note, that the acceleration of a neutron (assuming $q_{\rm n}=10^{-22}q_{e}$
and $E=10^6\,\rm V/m$) would be about $a=10^{-8}\,\rm m/s^2$. It is easy to show, that 
gravity forces in the same direction as the electric field can be realized on the order of $a=10^{-6}\,\rm m/s^2$,
hence two orders of magnitude larger and surely detectable. 
Considering (\ref{eqn:SystematicTilt}) however, the necessity to install and even more to move massive bodies of some $10^3\,\rm kg$ 
next to the setup appears to be challenging. But with the non-ebbing interest of neutrons' interaction with gravity \cite{Nesvizhevsky2006} 
one should keep an eye on every chance. 

We thank A.\,Franck from the JINR Dubna / Russia for stimulating discussions. This project is supported by the Carl-Zeiss-Foundation and is part of
the priority program 1491, 'Precision Experiments in Particle and Astrophysics with Cold and Ultracold Neutrons' of the Deutsche Forschungsgemeinschaft.

\bibliographystyle{unsrt}	
\bibliography{/Users/plonka/bibtex/myrefs}		





\end{document}